\documentclass[aps,
twocolumn, 
superscriptaddress,showpacs,pra,amssymb, amsmath]{revtex4}
\usepackage{amsmath,latexsym,amssymb}
\usepackage{graphicx}
\usepackage{amsmath}
\usepackage{latexsym}
\usepackage{amssymb}
\usepackage{bm}

\begin{document}

\title{Whether quantum mechanics can be almighty even in information science}

\author{Koji Nagata}
\affiliation{ 
Korea Advanced Institute of Science and Technology, Department of Physics}

\author{Tadao Nakamura}
\affiliation{Imperial College, London Department of Computing}
\affiliation{Stanford University, Computer Systems Laboratory}
\affiliation{Keio University
Science and Technology}

\date{\today}

\begin{abstract}
We discuss that there is a crucial contradiction within 
quantum mechanics.
We derive a proposition concerning a quantum expectation value 
under the assumption of the existence of the
directions in a spin-1/2 system. 
The quantum predictions within the formalism of von Neumann's 
projective measurement cannot coexist 
with the proposition concerning the existence of the
directions.
Therefore, we have to give up either the existence of the directions or 
the formalism of von Neumann's 
projective measurement.
Hence there is a crucial contradiction
within the Hilbert space formalism of 
the quantum theory.
This implies that there is no axiomatic system for the quantum theory.
This also reveals that we need new physical theories in order to 
explain the handing of raw experimental data.
We discuss that this crucial contradiction makes the quantum-theoretical formulation of Deutsch's algorithm
questionable.
\end{abstract}

\maketitle

As a famous physical theory, the quantum theory
(cf. \cite{Neumann,RPF,Redhead,Peres,JJ,NIELSEN_CHUANG}) 
gives accurate and at times 
remarkably accurate numerical predictions. 
Much experimental data 
fits to the quantum predictions 
for the past some 100 years.
The quantum theory also says 
new science with respect to information theory.
The science is called the quantum information theory \cite{NIELSEN_CHUANG}.
Therefore, the quantum theory gives us very useful another
theory in order to create a new information science and to explain the handing of raw experimental data.

As for the foundations of the quantum theory, Leggett-type 
nonlocal variables theory \cite{Leggett} 
is experimentally investigated \cite{1,2,3}.
The experiments report that the quantum theory 
does not accept Leggett-type nonlocal variables interpretation. 
As for the applications of the quantum theory, 
there are several attempts to use single-photon two-qubit
states for quantum computing. Oliveira {\it et al.} implement
Deutsch's algorithm \cite{DEUTSCH} with 
polarization and transverse spatial modes of
the electromagnetic field as qubits~\cite{Oliveira}. Single-photon
Bell states are prepared and measured~\cite{Kim2003}. Also
the decoherence-free implementation of Deutsch's algorithm is reported 
by using such
single-photon and by using two logical qubits~\cite{Mohseni2003}.
More recently, a one-way based experimental 
implementation of Deutsch's algorithm is reported~\cite{Tame}.

To date, the quantum theory seems to be a successful physical theory and 
it looks to have no problem in order to use it experimentally.
Several researches address \cite{Neumann} the 
mathematical formulation of the quantum theory.
It is desirable that the quantum theory is also mathematically 
successful because we predict unknown physical phenomena precisely.
Sometimes such predictions are effective 
in the field of elementary particle physics. 
We endure much time in order
to see the fact by using, for example, Large-scale accelerator.
Further, Rolf Landauer says that {\it Information is Physical}~\cite{NIELSEN_CHUANG}.
We cannot create any computer without physical phenomena.
This fact motivates us to investigate the Hilbert space formalism of 
the quantum theory.

Here we aim to discuss that 
there is a crucial contradiction within the Hilbert space formalism of 
the quantum theory.
We know that a theory means a set of propositions. Unfortunately, we have to abandon that the quantum theory satisfies consistency, which is necessary in order to have axiomatic system.
This implies that there is no axiomatic system for the quantum theory.
A theory $K$ may be said to be consistent 
if any proposition, $A\in K$, 
belonging to the theory $K$ and 
the negation of the proposition, $A^{\neg}$, are not derived,
simultaneously. Otherwise, the theory $K$ may be said to be contradictory.
Our discussion says that, surprisingly, the quantum theory is 
a contradictory physical theory in order to 
explain the handing of raw experimental data. 
Especially, we reexamine the quantum-theoretical 
formulation of Deutsch's algorithm \cite{DEUTSCH} as the earliest quantum computer.
We result in the fact that the formulation is questionable despite the fact that we indeed have raw experimental data.

Our discussion is very important. The reason is that our discussion 
reveals that we need new physical theories in order to 
explain raw data informationally, to create new information 
science, and to predict new unknown physical phenomena efficiently.
What are new physical theories? We cannot answer it at this stage.
However, we expect that our discussion in this thesis could contribute to 
creating new physical theories in order to explain the handing of raw experimental data, to create new information science, and to predict new unknown physical phenomena efficiently.

Our thesis is organized as follows.  
We derive 
a proposition concerning a quantum expectation value under the assumption of
the existence of the directions in a spin-1/2 system. 
The quantum predictions within the formalism of von Neumann's projective 
measurement (the results of measurements are $\pm 1$) cannot coexist with 
the proposition concerning the existence of the directions.
Therefore, there is a crucial contradiction 
in the set of propositions of 
the quantum theory
in a spin-1/2 system, viz.,
there is no axiomatic system for the quantum theory.
This crucial contradiction makes the quantum-theoretical formulation of Deutsch's algorithm questionable.
What we need is only one pure spin-1/2 state lying in the $x$-$y$ plane 
(a two-dimensional state).

Throughout this thesis, 
we confine ourselves to the two-dimensional (e.g., electron spin, photon 
polarizations, and so on) and the discrete eigenvalue case.
The number of settings of measuring apparatuses is two (two-setting model).
These assumptions are used in several experimental situations.

First, we discuss that there is a 
contradiction within quantum mechanics.
Assume a pure spin-$1/2$ state $\psi$ lying in the $x$-$y$ plane.
Let $\vec \sigma$ be $(\sigma_x,\sigma_y,\sigma_z)$, 
the vector of Pauli operators.
The measurements (observables) on a spin-1/2 state lying in the $x$-$y$ plane 
of $\vec n\cdot\vec\sigma$ are parameterized by 
a unit vector $\vec n$ 
(its direction along which the spin component is measured).
Here, $\cdot$ is the scalar product in ${\bf R}^{\rm 3}$. 

We have a quantum expectation value $E^k_{\rm QM},~k=1,2$ as
\begin{eqnarray}
E^k_{\rm QM}\equiv {\rm Tr}[\psi \vec n_k\cdot \vec \sigma],~k=1,2.\label{et}
\end{eqnarray}
We have 
$\vec x\equiv \vec x^{(1)}$, $\vec y\equiv \vec x^{(2)}$, and 
$\vec z\equiv \vec x^{(3)}$ which are the Cartesian axes 
relative to which spherical angles are measured.
Let us write the two unit vectors in the 
plane defined by $\vec x^{(1)}$ and $\vec x^{(2)}$ in the following way:
\begin{eqnarray}
\vec n_k=\cos\theta_k \vec x^{(1)}+\sin\theta_k \vec x^{(2)}.
\label{vector}
\end{eqnarray}
Here, the angle $\theta_k$ takes only two values:
$
\theta_1=0,~\theta_2=\frac{\pi}{2}
$.

We derive a necessary condition for
the quantum expectation value
for the system in a pure spin-1/2 state lying in the $x$-$y$ plane
given in (\ref{et}).
We derive the possible values of the scalar product 
$\sum_{k=1}^2 \left(E^k_{\rm QM}\times E^k_{\rm QM}\right)
\equiv \Vert  E_{\rm QM}\Vert^2$.
$E^k_{\rm QM}$ is the quantum expectation value given in (\ref{et}).
We see that $\Vert  E_{\rm QM}\Vert^2=
\langle \sigma_x\rangle^2+\langle\sigma_y\rangle^2$. 
We use decomposition (\ref{vector}).
We introduce simplified notations as
$
T_{i}=
{\rm Tr}[\psi \vec{x}^{(i)}\cdot \vec \sigma ]
$
and
$
(c^1_k, c^2_k,)=(\cos \theta_k,
\sin\theta_k).
$
Then, we have
\begin{eqnarray}
&&\Vert E_{\rm QM}\Vert^2=\sum_{k=1}^2
\left(\sum_{i=1}^2T_{i}
c^{i}_k\right)^2  =  
\sum_{i=1}^2T_{i}^2\leq 1,
\label{EEvalue}
\end{eqnarray}
where we use the orthogonality relation
$\sum_{k=1}^2 ~ c_k^{\alpha} c_k^{\beta}  =  \delta_{\alpha,\beta}$.
From a proposition of the quantum theory, the Bloch sphere (the directions) 
with the value of 
$\sum_{i=1}^2T_{i}^2$ is bounded as 
$\sum_{i=1}^2T_{i}^2\leq 1$.
The reason of the condition 
(\ref{EEvalue}) is the Bloch sphere
$
\sum_{i=1}^3 
({\rm Tr}[\psi \vec{x}^{(i)}\cdot \vec \sigma])^2\leq 1.
$
Thus we derive a proposition concerning a quantum expectation value 
under the assumption of the existence of the directions
(in a spin-1/2 system), that is, $\Vert E_{\rm QM}\Vert^2\le 1$.
It is worth noting here that
this inequality must be saturated if $\psi$ is 
a pure state lying in the $x$-$y$ plane. That is, 
$
\sum_{i=1}^2 
({\rm Tr}[\psi \vec{x}^{(i)}\cdot \vec \sigma])^2=1.
$
Hence we derive the following proposition 
concerning the existence of the directions
when the system is in a pure state lying in the $x$-$y$ plane
\begin{eqnarray}
\Vert E_{\rm QM}\Vert^2_{\rm max}= 1.\label{Bloch}
\end{eqnarray}
$\Vert E_{\rm QM}\Vert^2_{\rm max}$ is 
the maximal possible value of the scalar product.


On the other hand, let us assume von Neumann's projective measurement. 
In this case, the quantum expectation value in (\ref{et}), 
which is the average of the results of 
projective measurements, is given by
\begin{equation}
E^k_{\rm QM}=\lim_{m\rightarrow\infty}\frac{\sum_{l=1}^m r_l(\vec n_k)}{m}.
\label{avg}
\end{equation}
The possible values of the actually measured result $r_l(\vec n_k)$ are $\pm 1$ (in $\hbar/2$ unit). 
Same quantum expectation value is given by
\begin{eqnarray}
E^k_{\rm QM}=\lim_{m'\rightarrow\infty}\frac{\sum_{l'=1}^{m'} r_{l'}
(\vec n_k)}{m'},
\end{eqnarray}
because we only change the labels as $m\rightarrow m'$ and $l\rightarrow l'$.
Of course, the possible values of the actually 
measured result $r_{l'}(\vec n_k)$ are $\pm 1$ (in $\hbar/2$ unit). 
Thus, we have 
\begin{eqnarray}
\{l|l\in{\bf N}\wedge r_l(\vec n_k)=1\}
=
\{l'|l'\in{\bf N}\wedge r_{l'}(\vec n_k)=1\},
\end{eqnarray}
and
\begin{eqnarray}
\{l|l\in{\bf N}\wedge r_l(\vec n_k)=-1\}
=
\{l'|l'\in{\bf N}\wedge r_{l'}(\vec n_k)=-1\}.
\nonumber\\
\end{eqnarray}
By using these facts, we derive a necessary condition for
the quantum expectation value
for the system in a pure spin-1/2 state lying in the $x$-$y$ plane 
given in (\ref{avg}).
Again, we derive the possible values of the scalar product 
$\Vert E_{\rm QM}\Vert^2$
of the quantum expectation value, 
$E^k_{\rm QM}$ given in (\ref{avg}).
We have
\begin{eqnarray}
&&\Vert E_{\rm QM}\Vert^2\nonumber\\
&&=\sum_{k=1}^2
\left(\lim_{m\rightarrow\infty}\frac{\sum_{l=1}^m r_l(\vec n_k)}{m}
 \times 
 \lim_{m'\rightarrow\infty}\frac{\sum_{l'=1}^{m'} r_{l'}(\vec n_k)}{m'}
 \right)\nonumber\\
&&=\sum_{k=1}^2
\left(\lim_{m\rightarrow\infty}\frac{\sum_{l=1}^m }{m}\cdot
\lim_{m'\rightarrow\infty}\frac{\sum_{l'=1}^{m'} }{m'}
r_l(\vec n_k)
r_{l'}(\vec n_k)
 \right)
\nonumber\\
&&\leq\sum_{k=1}^2
\left(\lim_{m\rightarrow\infty}\frac{\sum_{l=1}^m }{m}\cdot
\lim_{m'\rightarrow\infty}\frac{\sum_{l'=1}^{m'} }{m'}
|r_l(\vec n_k)
r_{l'}(\vec n_k)|
 \right)
\nonumber\\
&&=\sum_{k=1}^2
\left(\lim_{m\rightarrow\infty}\frac{\sum_{l=1}^m }{m}\cdot
\lim_{m'\rightarrow\infty}\frac{\sum_{l'=1}^{m'} }{m'}
 \right)
 =2.
\label{integral}
\end{eqnarray}
Clearly, the above inequality can be saturated since, as we have said,
\begin{eqnarray}
\{l|l\in{\bf N}\wedge r_l(\vec n_k)=1\}
=
\{l'|l'\in{\bf N}\wedge r_{l'}(\vec n_k)=1\},
\end{eqnarray}
and
\begin{eqnarray}
\{l|l\in{\bf N}\wedge r_l(\vec n_k)=-1\}
=
\{l'|l'\in{\bf N}\wedge r_{l'}(\vec n_k)=-1\}.
\nonumber\\
\end{eqnarray}
Thus we derive a proposition concerning a quantum expectation value 
under the assumption that von Neumann's projective measurement is true
(in a spin-1/2 system), that is, $\Vert E_{\rm QM}\Vert^2\le 2$.
Hence we derive the following 
proposition concerning von Neumann's projective measurement
\begin{eqnarray}
\Vert E_{\rm QM}\Vert^2_{\rm max}= 2.\label{BSF}
\end{eqnarray}
Clearly, we cannot assign the truth value ``1'' 
for two propositions (\ref{Bloch}) (concerning 
the existence of the directions) 
and (\ref{BSF}) (concerning von Neumann's projective measurement), 
simultaneously,
when the system is in a pure state lying in the $x$-$y$ plane.
Therefore, we are in the contradiction when 
the system is in a pure state lying in the $x$-$y$ plane.

Next, we review Deutsch's algorithm 
along with Ref.~\cite{NIELSEN_CHUANG}.

Quantum parallelism is a fundamental feature of many quantum algorithms.
It allows quantum computers to evaluate the values of a function $f(x)$ for many different values of $x$ simultaneously.
Suppose $f:\{0,1\}\rightarrow \{0,1\}$ 
is a function with a one-bit domain and range.
A convenient way of computing this function on a quantum computer is to consider a two-qubit quantum computer which starts in the state $|x,y\rangle$.
With an appropriate sequence of logic gates it is possible to transform this state into $|x,y\oplus f(x)\rangle$, where $\oplus$ indicates addition modulo 2.
We give the transformation defined by 
the map $|x,y\rangle\rightarrow |x,y\oplus f(x)\rangle$ a name, $U_f$.

Deutsch's algorithm combines quantum parallelism with a property of quantum mechanics known as interference.
Let us use the Hadamard gate to prepare the first qubit $|0\rangle$
as the superposition $(|0\rangle+|1\rangle)/\sqrt{2}$, 
but let us prepare the second qubit as the superposition 
$(|0\rangle-|1\rangle)/\sqrt{2}$, using the Hadamard gate applied to 
the state $|1\rangle$.
The Hadamard gate is as 
$
H=\frac{1}{\sqrt{2}}(|0\rangle\langle 1|+|1\rangle\langle 0|+|0\rangle\langle 0|-|1\rangle\langle 1|).
$
Let us follow the states along to see what happens in this circuit.
The input state
\begin{eqnarray}
|\psi_0\rangle=|01\rangle\label{D1}
\end{eqnarray}
is sent through two Hadamard gates to give
\begin{eqnarray}
|\psi_1\rangle=\left[\frac{|0\rangle+|1\rangle}{\sqrt{2}}\right]
\left[\frac{|0\rangle-|1\rangle}{\sqrt{2}}\right].\label{D2}
\end{eqnarray}
A little thought shows that if we apply $U_f$ 
to the state $|x\rangle(|0\rangle-|1\rangle)/\sqrt{2}$ 
then we obtain the state $(-1)^{f(x)}|x\rangle(|0\rangle-|1\rangle)/\sqrt{2}$.
Applying $U_f$ to $|\psi_1\rangle$ therefore leaves us with one of two possibilities:
\begin{eqnarray}
|\psi_2\rangle=\left\{
\begin{array}{cl}
\displaystyle
\pm\left[\frac{|0\rangle+|1\rangle}{\sqrt{2}}\right]
\left[\frac{|0\rangle-|1\rangle}{\sqrt{2}}\right]
&\quad {\rm if}\ \ f(0)=f(1)\\
\\
\displaystyle
\pm\left[\frac{|0\rangle-|1\rangle}{\sqrt{2}}\right]
\left[\frac{|0\rangle-|1\rangle}{\sqrt{2}}\right]
&\quad {\rm if}\ \ f(0)\neq f(1).
\end{array} \right.\label{D3}
\end{eqnarray}
The final Hadamard gate on the first qubit thus gives us
\begin{eqnarray}
|\psi_3\rangle=\left\{
\begin{array}{cl}
\displaystyle
\pm|0\rangle
\left[\frac{|0\rangle-|1\rangle}{\sqrt{2}}\right]
&\quad {\rm if}\ \ f(0)=f(1)\\
\\
\displaystyle
\pm|1\rangle
\left[\frac{|0\rangle-|1\rangle}{\sqrt{2}}\right]
&\quad {\rm if}\ \ f(0)\neq f(1).
\end{array} \right.\label{D4}
\end{eqnarray}
Realizing that $f(0)\oplus f(1)$ is $0$ if $f(0)= f(1)$ and $1$ otherwise, 
we can rewrite this result concisely as
\begin{eqnarray}
|\psi_3\rangle=\pm|f(0)\oplus f(1)\rangle
\left[\frac{|0\rangle-|1\rangle}{\sqrt{2}}\right],
\end{eqnarray}
so by measuring the first qubit we may determine $f(0)\oplus f(1)$.
This is very interesting indeed: the quantum circuit has given us the ability to determine a global property of $f(x)$, namely $f(0)\oplus f(1)$, 
using only one evaluation of $f(x)$!
This is faster than is possible with a classical apparatus, which would require at least two evaluations.

In what follows, we discuss a problem of Deutsch's algorithm.
We see that the implementation of Deutsch's algorithm
is not possible if we give up either {\it observability} of a quantum state
or {\it controllability} of a quantum state.

We introduce the following quantum proposition concerning controllability:
\begin{eqnarray}
\langle 0|0\rangle=1,
\langle 1|1\rangle=1,
\langle 0|1\rangle=0,\ {\rm and}\
\langle 1|0\rangle=0.\label{proj}
\end{eqnarray}
We may consider the following non-quantum-theoretical proposition:
\begin{eqnarray}
\langle 0|0\rangle=-1,
\langle 1|1\rangle=-1,
\langle 0|1\rangle=0,\ {\rm and}\
\langle 1|0\rangle=0.\label{proj2}
\end{eqnarray}
The proposition (\ref{proj2}) implies
the validity of von Neumann's projective measurement (observability).
The proposition (\ref{proj2}) implies
\begin{eqnarray}
|\langle 0|0\rangle|^2=1,
|\langle 1|1\rangle|^2=1,
|\langle 0|1\rangle|^2=0,\ {\rm and}\
|\langle 1|0\rangle|^2=0.
\nonumber\\
\end{eqnarray}
However, the validity of von Neumann's projective measurement does not 
imply the proposition (\ref{proj2}).
We see that the proposition (\ref{proj}) is not equivalent to 
von Neumann's projective measurement (observability).
We see that we can assign the truth value ``1'' for von Neumann's projective measurement (observability) and we can 
assign the truth value ``0'' 
for the proposition (\ref{proj}) concerning controllability.

The proposition (\ref{proj}) implies that 
$\Vert E_{\rm QM}\Vert^2_{\rm max}
=\langle\sigma_x\rangle^2+\langle\sigma_y\rangle^2=1$ 
when the system is in a pure state lying in the $x$-$y$ plane.
The reason is as follows:
Assume a pure state lying in the $x$-$y$ plane as
$
|\psi\rangle=\frac{|0\rangle+e^{i\phi}|1\rangle}{\sqrt{2}}
$
where $\phi$ is a phase.
Let us write $\sigma_x=|0\rangle\langle 1|+|1\rangle\langle 0|$ and
$\sigma_y=-i|0\rangle\langle 1|+i|1\rangle\langle 0|$.
Then we have $\langle\psi|\sigma_x|\psi\rangle=\cos(\phi)$ and 
$\langle\psi|\sigma_y|\psi\rangle=\sin(\phi)$.
Therefore, we see 
$\langle\psi|\sigma_x|\psi\rangle^2+\langle\psi|\sigma_y|\psi\rangle^2=
\cos^2(\phi)+\sin^2(\phi)=1$.
We thus see the proposition (\ref{proj}) implies
that there are directions in the Hilbert space formalism of 
the quantum theory.

From the discussion presented in the previous, we see that
the quantum proposition (\ref{proj}) concerning controllability 
(the directions)
cannot coexist with
the validity of von Neumann's projective measurement 
(observability), which states
$\Vert E_{\rm QM}\Vert^2_{\rm max}=2$,
when the system is in a
pure state lying in the $x$-$y$ plane.

Deutsch's algorithm shows the importance of the ability of 
the Hadamard gate (controllability and the existence of the directions)
for quantum computation.
The ability of the Hadamard gate is valid only when we assign the 
truth value ``1'' for the proposition (\ref{proj}) (the directions).
We see that the quantum state $(|0\rangle\pm|1\rangle)/\sqrt{2}$
is a pure state lying in the $x$-$y$ plane.
We can assign the truth value ``1'' for the ability of the Hadamard gate 
(controllability and the existence of the directions)
\begin{eqnarray}
H\left(\frac{|0\rangle+|1\rangle}{\sqrt{2}}\right)=|0\rangle,~
H\left(\frac{|0\rangle-|1\rangle}{\sqrt{2}}\right)=|1\rangle
\end{eqnarray}
only when we assign the truth value ``1'' for the proposition 
(\ref{proj}) concerning controllability (directions) and
we give up the validity of 
von Neumann's projective measurement (observability).
The validity of the proposition (\ref{proj}) implies that $H^2=I$.
Thus applying $H$ 
twice to a state does nothing to it if we accept the proposition (\ref{proj}).
When we accept the proposition (\ref{proj}), we have 
\begin{eqnarray}
\frac{|0\rangle+|1\rangle}{\sqrt{2}}=H|0\rangle,~
\frac{|0\rangle-|1\rangle}{\sqrt{2}}=H|1\rangle.
\end{eqnarray}
We conclude that the step in which transforms 
the state $|\psi_0\rangle$ into the state $|\psi_1\rangle$, namely
the step saying from (\ref{D1}) to (\ref{D2}) is possible only when we assign the truth value ``1'' for the proposition 
(\ref{proj}) 
(concerning controllability and the existence of the directions) and 
we give up the validity of von Neumann's projective measurement 
(observability).
The step saying from (\ref{D3}) to (\ref{D4}) is also so.
Therefore we question what makes observability 
if we accept the ability of 
the Hadamard gate (controllability and the directions).
We also question what makes controllability
if we accept the validity of von Neumann's projective measurement 
(observability).

In conclusion, it may have been said that the quantum predictions within the formalism of von Neumann's 
projective measurement cannot coexist with 
the existence of the directions. 
These quantum-theoretical propositions have been contradicted each other.
Therefore there has been a crucial contradiction in 
the set of propositions of the quantum theory.
Hence there has been no 
informationally axiomatic system for the quantum theory.
Our discussion has been obtained in a quantum system which is in a pure 
spin-1/2 state lying in the $x$-$y$ plane.
We have reexamined the quantum-theoretical formulation of Deutsch's algorithm as the earliest quantum computer.
We have resulted in the fact that the formulation has been questionable despite the fact that we have indeed had raw experimental data.

What are new physical theories? We cannot answer it at this stage.
However, we expect that our discussion in this thesis could contribute to 
creating new physical theories in order to explain the handing of raw experimental data, to create new information science, and to predict new unknown physical phenomena efficiently.



\begin{acknowledgments}
We thank K. Niizeki and C. -L. Ren for valuable comments.
This work has been supported by Frontier Basic Research Programs at
KAIST and K.N. is supported by a BK21 research grant.
\end{acknowledgments}



\end{document}